\documentclass[seceq]{ptptex}

\usepackage{graphicx}
\usepackage{wrapft}

\newcommand{\1}{\mbox{1}\hspace{-0.25em}\mbox{l}} 

\title{
Running Boundary Condition%
}

\author{
Satoshi \textsc{Ohya},$^{1,}$\footnote{E-mail: \texttt{satoshi.ohya@pi.infn.it}}
Makoto \textsc{Sakamoto}$^{2,}$\footnote{E-mail: \texttt{dragon@kobe-u.ac.jp}}
and
Motoi \textsc{Tachibana}$^{3,}$\footnote{E-mail: \texttt{motoi@cc.saga-u.ac.jp}}%
}

\inst{
${}^{1}$INFN Sezione di Pisa, Largo Bruno Pontecorvo 3, 56127 Pisa, Italy\\
${}^{2}$Department of Physics, Kobe University, Kobe 657-8501, Japan\\
${}^{3}$Department of Physics, Saga University, Saga 840-8502, Japan
}

\abst{
In this paper we argue that boundary condition may run with energy scale.
As an illustrative example, we consider one-dimensional quantum mechanics for a spinless particle that freely propagates in the bulk yet interacts only at the origin.
In this setting we find the renormalization group flow of $U(2)$ family of boundary conditions exactly.
We show that the well-known scale-independent subfamily of boundary conditions are realized as fixed points.
We also discuss the duality between two distinct boundary conditions from the renormalization group point of view.
Generalizations to conformal mechanics and quantum graph are also discussed.
}

\begin{document}

\maketitle

\section{Introduction} \label{sec:intro}
Boundary condition has become more and more relevant not only for condensed matter physics but also for high-energy physics.
Variety of boundary conditions has provided various interesting phenomena, such as supersymmetry,\cite{Fulop:1999pf,Uchino:2002xb,Nagasawa:2002un,Uchino:2003kh,Nagasawa:2003tw,Nagasawa:2005kv} duality\cite{Cheon:1998iy,Tsutsui:2000,Cheon:2000tq}, anholonomy,\cite{Tsutsui:2000,Cheon:2000tq} and conformal to nonconformal phase transition.\cite{Kaplan:2009kr}
As is evident in the context of impurity problems, nontrivial boundary condition imposed on wavefunctions or quantum fields can be regarded as the presence of point interaction, or interaction of zero range:
For example, one-dimensional quantum mechanics with a delta function potential described by the Hamiltonian $H = -\mathrm{d}^{2}/\mathrm{d}x^{2} + 2g\delta(x)$ is equivalent to the system described by the bulk Hamiltonian $H_{\text{bulk}} = -\mathrm{d}^{2}/\mathrm{d}x^{2}$ plus the boundary condition at the origin $\psi(0_{+}) = \psi(0_{-})$ and $\psi^{\prime}(0_{+}) - \psi^{\prime}(0_{-}) = g(\psi(0_{+}) + \psi(0_{-}))$. (Here prime (${}^{\prime}$) indicates the derivative with respect to $x$.)
Another related example is the presence of zero-thickness brane in extra dimensional models.
In five-dimensional model with a single extra dimension with extended defects or zero-thickness branes, for example, operators such as mass terms localized to the position of defects are often introduced because at such points translational invariance is generically broken.
These mass terms, however, can be described by boundary conditions of quantum fields just like the delta function potential in quantum mechanics.
In this sense, an imposition of nontrivial boundary condition to wavefunctions/quantum fields can be viewed as to introduce a nontrivial point interaction at a certain spatial point.

In the actual real world, however, there is no point-like structure at all energy scale:
Point impurity may be an atom, which has its own actual size and is spatially extended but it can be approximated by a point when one limits to consider energy scale small enough compared to the actual size of the atom;
Zero-thickness brane may be a nonzero-thickness brane, which should be realized as solitonic configuration of string theory and has its own spatial extent but it could be approximated to zero-thickness when one limits to consider energy scale much below the thickness of brane.
Point interactions and hence corresponding boundary conditions must be effective descriptions of actual finite range interactions valid for the energy scale much below the size of localized interactions.

The concept behind the above paragraph is the following naive expectation; that is, any short-ranged interaction could be approximated by a point interaction in the long-wavelength limit.
In order to get detailed information about what the localized interaction is, we need to use a probe particle whose de Broglie wavelength is shorter than the size of the localized interaction.
In other words, the longer the probe particle's wavelength is, the less information we can get about the short-ranged interaction.
This naive consideration leads to the following elementary question:
\textit{``Do there exist any universality classes of short-ranged interactions whose long-wavelength limits appear to be the same?''}, or, equivalently, \textit{``Do there exist any universality classes of boundary conditions whose low-energy limit reduces to the same?''}
In this paper we would like to try to argue this by investigating the renormalization group (RG) flow of boundary conditions.

As the simplest yet nontrivial setting, however, in this paper we will concentrate ourselves to one-particle quantum mechanics in one spatial dimension.\footnote{In the context of two-dimensional quantum field theory on a manifold with boundaries, RG flow which interpolates distinct scale-independent boundary conditions (or conformal boundary states) has been studied under the field of boundary conformal field theory; see for short review Ref.~\citen{Graham:2000si} and references therein.}
As we will see in the rest of the paper, this simple setting allows us to derive the RG flow of boundary conditions exactly.

To begin with, let us imagine one-dimensional quantum mechanics for a single spinless particle on $\mathbb{R}$ in the presence of a single localized potential centered at the origin, whose spatial extent is characterized by a length scale $a$.
In the long-wavelength limit $\lambda \gg a$ with $\lambda$ being the de Broglie wavelength of a probe particle, any localized potential could be approximated by a point interaction at the origin.
In this limit where $a$ plays a role of physical cutoff, a particle we consider would freely propagate in the bulk yet interact only at the origin.
The time-independent Schr\"odinger equation describing this situation must be as follows:
\begin{align}
H_{\text{bulk}}\psi(x)
&= 	E\psi(x), \quad
x\neq 0, 	\label{eq:1.1}
\end{align}
where the bulk Hamiltonian $H_{\text{bulk}}$ is given by
\begin{align}
H_{\text{bulk}}
&= 	- \frac{\mathrm{d}^{2}}{\mathrm{d}x^{2}}. \label{eq:1.2}
\end{align}
(In this paper we will work in the units where $\hbar = 2m = 1$.)
It is known that allowed point interactions in one-dimensional quantum mechanics are all described by the boundary conditions consistent with the self-adjointness of the Hamiltonian operator $H_{\text{bulk}}$.\cite{Cheon:2000tq}
In order to make this paper self-contained, let us recall the argument given in Ref.~\citen{Cheon:2000tq} with slight modifications for the purpose of this paper.
Since the Hamiltonian must be a generator of unitary time evolution operator, its self-adjointness indicates the conservation of probability in the whole system, especially even at the origin.
Thus we see that our requirement for the self-adjointness of $H_{\text{bulk}}$ is just equivalent to the conservation of probability current density at the origin:
\begin{align}
j(0_{+})
&= 	j(0_{-}), \label{eq:1.3}
\end{align}
where the probability current density is defined by
\begin{align}
j(x)
&:= 	-i\left[
	\psi^{\prime\ast}(x)\psi(x) - \psi^{\ast}(x)\psi^{\prime}(x)
	\right], \label{eq:1.4}
\end{align}
with $\psi$ being a wavefunction on $\mathbb{R}$.
Again prime (${}^{\prime}$) indicates the derivative with respect to $x$.
For the following discussions it is convenient to introduce the 2-component column vector $\Vec{\Psi}(x)$ and its derivative $\Vec{\Psi}^{\prime}(x)$ as
\begin{subequations}
\begin{align}
\Vec{\Psi}(x)
&:= 	\left(
	\psi(x), \psi(-x)
	\right)^{T}, \label{eq:1.5a}\\
\Vec{\Psi}^{\prime}(x)
&:= 	\left(
	\psi^{\prime}(x), -\psi^{\prime}(-x)
	\right)^{T}, \label{eq:1.5b}
\end{align}
\end{subequations}
where $T$ stands for the transposition of matrix.
Using these vectors the requirement \eqref{eq:1.3} can be rewritten as $\Vec{\Psi}^{\dagger}(0_{+})\cdot\Vec{\Psi}^{\prime}(0_{+}) = \Vec{\Psi}^{\prime\dagger}(0_{+})\cdot\Vec{\Psi}(0_{+})$, or, equivalently,
\begin{align}
\bigl|
\Vec{\Psi}(0_{+}) - iL_{0}\Vec{\Psi}^{\prime}(0_{+})
\bigr|^{2}
&= 	\bigl|
	\Vec{\Psi}(0_{+}) + iL_{0}\Vec{\Psi}^{\prime}(0_{+})
	\bigr|^{2}, \label{eq:1.6}
\end{align}
where $L_{0}$ is an \textit{arbitrary} non-vanishing real length scale, which is just introduced to adjust the length dimension of the equation \eqref{eq:1.6}.
Since our bulk Hamiltonian is characterized by lack of any scale parameters, we immediately see that any dimensionful quantities, such as momentum or energy of a particle, must be scaled by this arbitrary parameter.
As we will see in \S \ref{sec:RG}, $L_{0}$ turns out to play a role of a renormalization scale if we require any physical quantities such as scattering amplitudes or bound state energies should not depend on the choice of $L_{0}$.
The dependence of the theory on this scale parameter will be described by the RG.

Equation \eqref{eq:1.6} shows that the squared length of the 2-component complex column vector $\Vec{\Psi}(0_{+}) - iL_{0}\Vec{\Psi}^{\prime}(0_{+})$ is equal to that of $\Vec{\Psi}(0_{+}) + iL_{0}\Vec{\Psi}^{\prime}(0_{+})$, which implies that these two vectors must be related by a two-dimensional unitary transformation.
Thus we can write
\begin{align}
\Vec{\Psi}(0_{+}) - iL_{0}\Vec{\Psi}^{\prime}(0_{+})
&= 	U
	\bigl[
	\Vec{\Psi}(0_{+}) + iL_{0}\Vec{\Psi}^{\prime}(0_{+})
	\bigr], \quad
	U \in U(2). \label{eq:1.7}
\end{align}
This is the $U(2)$ family of boundary conditions that describe all possible point interactions in one-dimensional quantum mechanics.\cite{Cheon:2000tq}
Any point interaction will be specified by a certain unitary matrix $U \in U(2)$.
In this sense we can say that in an appropriate long-wavelength limit, the theory space of one-dimensional quantum mechanics for a particle with a single localized potential is equivalent to the parameter space of two-dimensional unitary group $U(2)$.

For the following discussions it is suitable to parameterize the matrix $U$ into the following spectral decomposition form:
\begin{align}
U
&= 	\mathrm{e}^{i\alpha_{+}}P_{+} + \mathrm{e}^{i\alpha_{-}}P_{-}, \quad
P_{\pm}
:= 	\frac{\1 \pm \Vec{e}\cdot\Vec{\sigma}}{2}, \label{eq:1.8}
\end{align}
where $0\leq\alpha_{\pm}<2\pi$ and $\Vec{e} = (e_{x}, e_{y}, e_{z})^{T}$ is a real unit vector satisfying $e_{x}^{2} + e_{y}^{2} + e_{z}^{2} = 1$.
$P_{\pm}$ are the hermitian projection operators fulfilling $P_{+} + P_{-} = \1$, $(P_{\pm})^{2} = P_{\pm}$, $P_{\pm}P_{\mp} = 0$ and $P_{\pm}^{\dagger} = P_{\pm}$.
By substituting \eqref{eq:1.8} into \eqref{eq:1.7}, the boundary condition boils down to the following two independent equations:
\begin{align}
P_{\pm}\bigl[\Vec{\Psi}(0_{+}) + L_{\pm}\Vec{\Psi}^{\prime}(0_{+})\bigr]
&= 	\Vec{0}, \label{eq:1.9}
\end{align}
where
\begin{align}
L_{\pm}
&:= 	L_{0}\cot(\alpha_{\pm}/2). \label{eq:1.10}
\end{align}
Notice that when $\alpha_{\pm} = 0$ or $\pi$ the scale parameters $L_{\pm}$ drop out from \eqref{eq:1.9}:
\begin{subequations}
\begin{align}
P_{\pm}\Vec{\Psi}^{\prime}(0_{+})
&= 	\Vec{0},
\quad\text{for~} \alpha_{\pm} = 0, \label{eq:1.11a}\\
P_{\pm}\Vec{\Psi}(0_{+})
&= 	\Vec{0},
\quad\text{for~} \alpha_{\pm} = \pi. \label{eq:1.11b}
\end{align}
\end{subequations}

The rest of this paper is organized as follows.
In \S \ref{sec:S-matrix} we will derive the one-particle scattering matrix (S-matrix) and bound state energies exactly.
In \S \ref{sec:RG} we will derive the exact RG flow of boundary conditions by using the exact S-matrix.
We will see that the scale-independent boundary conditions \eqref{eq:1.11a} and \eqref{eq:1.11b} correspond to the boundary conditions realized at the fixed points of RG flow.
In \S \ref{sec:duality} we will briefly discuss the duality between two distinct boundary conditions from the RG point of view.
Section \ref{sec:conclusions} is devoted to conclusions and discussions.

\section{S-matrix and bound state energy} \label{sec:S-matrix}
In this section we solve the Schr\"odinger equation \eqref{eq:1.1} with the boundary conditions \eqref{eq:1.9} and then derive the exact S-matrix and bound state energies.

\begin{figure}[t]
\centerline{\includegraphics{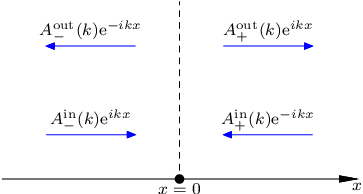}}
\caption{One-particle scattering from a point defect.}
\label{fig:2}
\end{figure}

The general solution to the Schr\"odinger equation \eqref{eq:1.1} for positive energy $E>0$ is the linear combination of the plane waves  
\begin{equation}
\psi(x; k)
= 	\begin{cases}
	\displaystyle A_{+}^{\text{in}}(k)\mathrm{e}^{-ikx} + A_{+}^{\text{out}}(k)\mathrm{e}^{ikx},
	& \text{for~} x>0, \\
	\displaystyle A_{-}^{\text{in}}(k)\mathrm{e}^{ikx} + A_{-}^{\text{out}}(k)\mathrm{e}^{-ikx},
	& \text{for~} x<0, \label{eq:2.1}
	\end{cases}
\end{equation}
where $k := \sqrt{E}>0$.
Note that the coefficients $A_{\pm}^{\text{in}}(k)$ and $A_{\pm}^{\text{out}}(k)$ may depend on $k$.
The superscripts `in' and `out' mean the incoming waves towards the origin and the outgoing waves against the origin, respectively (see Fig.~\ref{fig:2}).
The 2-component vectors \eqref{eq:1.5a} and \eqref{eq:1.5b} at the origin then become
\begin{subequations}
\begin{align}
\Vec{\Psi}(0_{+})
&= 	\Vec{A}^{\text{in}}(k) + \Vec{A}^{\text{out}}(k), \label{eq:2.2a}\\
\Vec{\Psi}^{\prime}(0_{+})
&= 	ik\bigl[- \Vec{A}^{\text{in}}(k) + \Vec{A}^{\text{out}}(k)\bigr], \label{eq:2.2b}
\end{align}
\end{subequations}
where $\Vec{A}^{\text{in}}(k)$ and $\Vec{A}^{\text{out}}(k)$ are $2$-component column vectors defined as
\begin{subequations}
\begin{align}
\Vec{A}^{\text{in}}(k)
&:= 	\bigl(A_{+}^{\text{in}}(k), A_{-}^{\text{in}}(k)\bigr)^{T}, \label{eq:2.3a}\\
\Vec{A}^{\text{out}}(k)
&:= 	\bigl(A_{+}^{\text{out}}(k), A_{-}^{\text{out}}(k)\bigr)^{T}. \label{eq:2.3b}
\end{align}
\end{subequations}
Substituting these into the boundary conditions \eqref{eq:1.9} we get
\begin{align}
\Vec{0}
&= 	P_{\pm}\bigl[\Vec{\Psi}(0_{+}) + L_{\pm}\Vec{\Psi}^{\prime}(0_{+})\bigr]
= 	(1+ikL_{\pm})P_{\pm}
	\left[\Vec{A}^{\text{out}}(k) - \frac{ikL_{\pm} - 1}{ikL_{\pm} + 1}\Vec{A}^{\text{in}}(k)\right]. \label{eq:2.4}
\end{align}
Since the factor $(1+ikL_{\pm})$ cannot be zero for real $k>0$, we get the following two independent equations:
\begin{align}
P_{\pm}
\left[\Vec{A}^{\text{out}}(k) - \frac{ikL_{\pm} - 1}{ikL_{\pm} + 1}\Vec{A}^{\text{in}}(k)\right]
&= 	\Vec{0}. \label{eq:2.5}
\end{align}
Since these two independent equations are orthogonal to each other, they can be combined into the following form:
\begin{align}
\Vec{0}
&= 	\sum_{j=\pm}
	P_{j}
	\left[\Vec{A}^{\text{out}}(k) - \frac{ikL_{j} - 1}{ikL_{j} + 1}\Vec{A}^{\text{in}}(k)\right]
= 	\Vec{A}^{\text{out}}(k) - \sum_{j=\pm}\frac{ikL_{j} - 1}{ikL_{j} + 1}P_{j}\Vec{A}^{\text{in}}(k), \label{eq:2.6}
\end{align}
where the second equality follows from $\sum_{j=\pm}P_{j} = \1$.
Thus,
\begin{align}
\Vec{A}^{\text{out}}(k)
&= 	S(k)\Vec{A}^{\text{in}}(k), \label{eq:2.7}
\end{align}
where $S(k)$ is a $2\times 2$ matrix defined as
\begin{align}
S(k)
&:= 	\sum_{j=\pm}\frac{ikL_{j} - 1}{ikL_{j} + 1}P_{j}
= 	\1 - 2\sum_{j=\pm}\frac{1}{ikL_{j} + 1}P_{j}. \label{eq:2.8}
\end{align}
Obviously the matrix $S(k)$ is unitary, $S^{\dagger}(k)S(k) = S(k)S^{\dagger}(k) = \1$, and has the modulus unity eigenvalues $(ikL_{\pm} - 1)/(ikL_{\pm} + 1)$.

Equation \eqref{eq:2.7} shows that the matrix $S(k)$ plays a role of an evolution map between the ``in-state'' $\Vec{A}^{\text{in}}(k)$ and the ``out-state'' $\Vec{A}^{\text{out}}(k)$.
The $ij$-component of the matrix $S(k)$ is nothing but the transition amplitude for a particle traveling from $\mathbb{R}_{j}$ to $\mathbb{R}_{i}$, where $i,j = +$ or $-$.
Thus, the diagonal element $S_{++}$ ($S_{--}$) should be interpreted as the reflection coefficient $R_{+}(k)$ ($R_{-}(k)$) for a particle of momentum $k$ on the positive (negative) half-line $\mathbb{R}_{+}$ ($\mathbb{R}_{-}$).
Similarly, the off-diagonal elements $S_{\pm\mp}$ should be interpreted as the transmission coefficients $T_{\pm}(k)$ for a particle incoming from $\mathbb{R}_{\mp}$ and scattered to $\mathbb{R}_{\pm}$.
Hence we interpret $S(k)$ as the one-particle S-matrix and write
\begin{align}
S(k)
&= 	\begin{pmatrix}
	R_{+}(k) 		& T_{+}(k) \\
	T_{-}(k) 		& R_{-}(k)
	\end{pmatrix}. \label{eq:2.9}
\end{align}
It should be emphasized that this S-matrix is \textit{exact}.

Let us next discuss the negative energy ($E<0$) state.
The negative energy or bound state solution is obtained by just replacing $k$ to $i\kappa$ ($\kappa>0$) in Eq.~\eqref{eq:2.1}.
The square integrability of the wavefunction on $\mathbb{R}$ requires $A_{\pm}^{\text{in}}(i\kappa) = 0$ and hence from Eq.~\eqref{eq:2.4} the nontrivial bound state solutions could exist if and only if $\kappa = 1/L_{\pm}$ with $L_{\pm}>0$, in which case the bound state wavefunctions are given by
\begin{align}
\psi_{B}^{\pm}(x)
&\propto
	\exp\left(-\frac{|x|}{L_{\pm}}\right), \quad
	-\infty < x < \infty. \label{eq:2.10}
\end{align}
The corresponding energy eigenvalues are
\begin{align}
E_{B}^{\pm}
&= 	-\frac{1}{L_{\pm}^{2}}. \label{eq:2.11}
\end{align}
Without any loss of generality, we can assume that $L_{0}>0$.
With this assumption the number of bound states appeared in the spectrum is classified as follows:
\begin{alignat}{5}
&\text{(i)}&\quad&\text{zero bound state}&\quad
&\text{for} \quad (\alpha_{+}, \alpha_{-}) = (0, 0), (0, \pi), (\pi, 0), (\pi, \pi),& \nonumber\\
&              &          &                                        &\quad
&\phantom{\text{for}} \quad (\pi, \pi)<(\alpha_{+}, \alpha_{-})<(2\pi, 2\pi);& \nonumber\\
&\text{(ii)}&\quad&\text{two bound states}&\quad
&\text{for} \quad (0, 0)<(\alpha_{+}, \alpha_{-})<(\pi, \pi);& \nonumber\\
&\text{(iii)}&\quad&\text{a single bound state}&\quad
&\text{otherwise}.& \nonumber
\end{alignat}

Whether the bound states exist or not can also be explained from the scattering theory point of view.
It follows from the boundary conditions \eqref{eq:2.4} that at $k = i/L_{\pm}$, which is a simple pole in the S-matrix, the general solution \eqref{eq:2.1} with $\Vec{A}^{\text{in}}(i/L_{\pm}) = \Vec{0}$ behaves as $\psi(x; k = \tfrac{i}{L_{\pm}}) \propto \exp(-\tfrac{|x|}{L_{\pm}})$.
When $k$ is in the upper half $k$-plane (i.e. $L_{\pm}>0$), the wave function at the pole exhibits an exponentially damping behavior asymptotically and hence is normalizable.
This shows that if a simple pole in the S-matrix lies on the positive imaginary $k$-axis, it corresponds to a bound state.
When $k$ is in the lower half $k$-plane (i.e. $L_{\pm}<0$), the wave function has an exponentially growing behavior asymptotically and thus cannot be normalizable.
Such a non-normalizable solution is usually referred to as  an \textit{antibound} (or \textit{virtual}) \textit{state}.
As we will see in \S \ref{sec:duality}, under the duality a normalizable bound state transforms into a non-normalizable antibound state and vice versa.

\section{Exact RG flow of boundary conditions} \label{sec:RG}
In this section we study the RG flow of $U(2)$ family of boundary conditions.
To this end, let us first investigate the ultraviolet (UV) and infrared (IR) behaviors of the S-matrix and give an observation to boundary conditions realized at the fixed points of RG flow.

When $\alpha_{\pm} \neq 0$ or $\pi$ (i.e. $L_{\pm}\neq \infty$ or $0$), the S-matrix $S(k)$ flows into the unit matrix $\1$ in the UV regime ($k\to\infty$), while in the IR regime ($k\to0$) it flows into $-\1$.
Since $S(k) = \pm\1$ means $U = \pm\1$, we see that in this case the point interactions flow into the infinite walls described by the Dirichlet and Neumann boundary conditions in the IR and UV limits, respectively; see Eq.~\eqref{eq:1.7}.
Note further that the diagonal S-matrix means the transmissionless point interactions, in this case particles cannot penetrate through the origin both in the UV and IR limits.
This is physically equivalent to the situation where a single line splits into two disconnected half lines in the UV and IR limits.

The above situation will be changed when $\alpha_{\pm} = 0$ or $\pi$.
Let us first consider the case where $\alpha_{+} = 0$ and $\alpha_{-}\neq 0, \pi$.
In this case the S-matrix flows as $S(k) \stackrel{k\to\infty}{\to} \1$ and $S(k) \stackrel{k\to0}{\to} P_{+} - P_{-} = \Vec{e}\cdot\Vec{\sigma}$, which means that while in the IR regime particles can penetrate through the origin, it is impossible in the UV regime.
As a next example let us consider the case where $\alpha_{+} = \pi$ and $\alpha_{-}\neq 0, \pi$.
In this case the S-matrix flows as $S(k) \stackrel{k\to\infty}{\to} -P_{+}+P_{-} = - \Vec{e}\cdot\Vec{\sigma}$ and $S(k) \stackrel{k\to0}{\to} -\1$.
As contrast to the previous case, in this case particles can penetrate through the origin in the UV regime but it is impossible in the IR regime.
All of the different behaviors of the S-matrix are summarized in Table \ref{tab:1}.
Different flows of the S-matrix in the IR limit implies that there exist nontrivial fixed points in the $U(2)$ parameter space (theory space).
In what follows we will confirm that this observation is indeed true.
\begin{table}[t]
\begin{center}
\caption{Flow of the S-matrix $S(k)$.}
\label{tab:1}
\begin{tabular}{ll | ll}
\hline
$\alpha_{+}$ 			& $\alpha_{-}$ 			& UV ($k\to\infty$) 								& IR ($k\to0$) \\
\hline
$\alpha_{+}\neq0,\pi$ 	& $\alpha_{-}\neq0,\pi$ 	& $S(k)\to \1$ 									& $S(k)\to -\1$ \\
$\alpha_{+} = 0$ 		& $\alpha_{-}\neq 0, \pi$ 	& $S(k)\to \1$ 									& $S(k)\to P_{+} - P_{-} = \Vec{e}\cdot\Vec{\sigma}$ \\
$\alpha_{+} \neq 0, \pi$ 	& $\alpha_{-} = 0$ 		& $S(k)\to \1$ 									& $S(k)\to -P_{+} + P_{-} = -\Vec{e}\cdot\Vec{\sigma}$\\
$\alpha_{+} = \pi$ 		& $\alpha_{-}\neq 0,\pi$ 	& $S(k)\to -P_{+} + P_{-} = -\Vec{e}\cdot\Vec{\sigma}$ 	& $S(k)\to -\1$ \\
$\alpha_{+} \neq 0,\pi$ 	& $\alpha_{-} = \pi$ 		& $S(k)\to P_{+} - P_{-} = \Vec{e}\cdot\Vec{\sigma}$ 		& $S(k)\to -\1$ \\
\hline
$\alpha_{+} = 0$ 		& $\alpha_{-} = 0$ 		& $S(k) = \1$ 									& $S(k) = \1$ \\
$\alpha_{+} = \pi$ 		& $\alpha_{-} = \pi$ 		& $S(k) = -\1$ 									& $S(k) = -\1$ \\
$\alpha_{+} = 0$ 		& $\alpha_{-} = \pi$ 		& $S(k) = \Vec{e}\cdot\Vec{\sigma}$ 					& $S(k) = \Vec{e}\cdot\Vec{\sigma}$ \\
$\alpha_{+} = \pi$ 		& $\alpha_{-} = 0$ 		& $S(k) = -\Vec{e}\cdot\Vec{\sigma}$ 				& $S(k) = -\Vec{e}\cdot\Vec{\sigma}$ \\
\hline
\end{tabular}
\end{center}
\end{table}

\subsection[Exact beta-function]{Exact $\beta$-function} \label{sec:beta-function}
\begin{figure}[t]
\begin{center}
\begin{tabular}{cc}
\begin{minipage}{6.6cm}
\begin{center}
\includegraphics{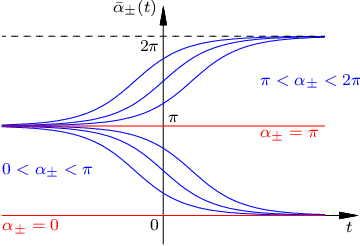}
\caption{Running eigenphase ${\bar \alpha}_{\pm}(t)$ with initial values $\alpha_{\pm} = 0, \frac{\pi}{4}, \frac{\pi}{2}, \frac{3\pi}{4}, \pi, \frac{5\pi}{4}, \frac{3\pi}{2}, \frac{7\pi}{4}$.}
\label{fig:Running}
\end{center}
\end{minipage}
&
\begin{minipage}{6.6cm}
\begin{center}
\includegraphics{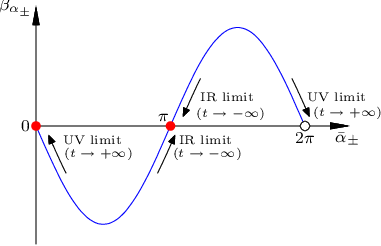}
\caption{Exact $\beta$-function for ${\bar \alpha}_{\pm}$. $\Bar{\alpha}_{\pm} = 0$ ($\pi$) is the UV(IR) fixed point.}
\label{fig:Beta}
\end{center}
\end{minipage}
\end{tabular}
\end{center}
\end{figure}

As noted before, $L_{0}$ is an arbitrary reference scale so that the physical quantities, such as an S-matrix element or a bound state energy, must be independent of the choice of $L_{0}$.
The lack of dependence of $L_{0}$ can be expressed as an invariance of the theory under the RG transformation
\begin{align}
R_{t}: L_{0}\mapsto {\bar L}(t) := L_{0}\mathrm{e}^{-t}, \quad -\infty<t<\infty. \label{eq:3.1}
\end{align}
Any change of $L_{0}$ must be equivalent to changes in the $U(2)$ parameters $g_{i} = \{\alpha_{\pm}, e_{i}\}$.
This requirement is expressed as
\begin{subequations}
\begin{align}
S(k; g_{i}, L_{0})
&= 	S(k; {\bar g}_{i}(t), {\bar L}(t)), \label{eq:3.2a}\\
E_{B}^{\pm}(\alpha_{\pm}, L_{0})
&= 	E_{B}^{\pm}({\bar \alpha}_{\pm}(t), {\bar L}(t)), \label{eq:3.2b}
\end{align}
\end{subequations}
where ${\bar g}_{i}(t) = \{{\bar \alpha}_{\pm}(t), {\bar e}_{i}(t)\}$ are the running $U(2)$ parameters, which will be determined by the following two equivalent ways:
\begin{enumerate}
\item
Before embarking on a standard RG approach, it is wise to consider first the S-matrix evaluated at momentum $k\mathrm{e}^{t}$.
Dimensional analysis and the invariance of S-matrix under the RG transformation allow us to relate it to the S-matrix at momentum $k$:
\begin{align}
S(k\mathrm{e}^{t}; g_{i}, L_{0})
&= 	S(k; g_{i}, L_{0}\mathrm{e}^{t})
= 	S(k; {\bar g}_{i}(t), (L_{0}\mathrm{e}^{t})\mathrm{e}^{-t})
= 	S(k; {\bar g}_{i}(t), L_{0}), \label{eq:3.3}
\end{align}
where the first equality follows from the dimensional analysis: the S-matrix is a dimensionless quantity and hence its momentum dependence must be encoded with the combination $kL_{\pm}$.
The second equality, on the other hand, follows from Eq.~\eqref{eq:3.2a}.
Since momentum $k$ appears only in the combination $kL_{\pm} = kL_{0}\cot(\alpha_{\pm}/2)$, the rescaling of $k$ must be adjusted by the running of $\alpha_{\pm}$:
\begin{align}
kL_{\pm}
&\stackrel{k\mapsto k\mathrm{e}^{t}}{\mapsto}
	(k\mathrm{e}^{t})L_{\pm}
= 	k(L_{\pm}\mathrm{e}^{t})
= 	k\left(L_{0}\mathrm{e}^{t}\cot\frac{\alpha_{\pm}}{2}\right)
= 	k\left(L_{0}\cot\frac{{\bar \alpha}_{\pm}(t)}{2}\right), \label{eq:3.4}
\end{align}
where the last equality follows from the requirement \eqref{eq:3.2a}.
Thus, in order for the invariance of the theory under the RG transformation we must have
\begin{align}
\cot\frac{{\bar \alpha}_{\pm}(t)}{2}
&= 	\mathrm{e}^{t}\cot\frac{\alpha_{\pm}}{2}, \label{eq:3.5}
\end{align}
from which we obtain
\begin{align}
{\bar \alpha}_{\pm}(t)
&= 	2\arctan\left(\mathrm{e}^{-t}\tan\frac{\alpha_{\pm}}{2}\right)
= 	\frac{1}{i}\log\left(\frac{1 + i\mathrm{e}^{-t}\tan(\alpha_{\pm}/2)}{1 - i\mathrm{e}^{-t}\tan(\alpha_{\pm}/2)}\right). \label{eq:3.6}
\end{align}
The running of $e_{i}$, on the other hand, must be trivial; that is, it must be exactly marginal
\begin{align}
{\bar e}_{i}(t)
&= 	e_{i}. \label{eq:3.7}
\end{align}
The running of ${\bar \alpha}_{\pm}(t)$ for several initial values $\alpha_{\pm}$ is depicted in Fig.~\ref{fig:Running}.

\item
Let us next rederive the above results \eqref{eq:3.5} and \eqref{eq:3.7} by using the RG techniques.
Since $S(k; g_{i}, L_{0})$ dose not have $t$ in any way, from Eq.~\eqref{eq:3.2a} we must have
\begin{align}
\frac{\partial}{\partial t}S(k; g_{i}, L_{0})\biggl|_{g_{i}, L_{0}}
&= 	0
= 	\frac{\partial}{\partial t}
	S(k; {\bar g}_{i}(t), {\bar L}(t))\biggl|_{g_{i}, L_{0}}. \label{eq:3.8}
\end{align}
The first equality is trivial, but the the second one leads to the following homogeneous RG equation:
\begin{align}
\left(
- {\bar L}\frac{\partial}{\partial {\bar L}}
+ \sum_{{\bar g}_{i} = {\bar \alpha}_{\pm}, {\bar e}_{i}}
\beta_{g_{i}}({\bar g}_{i}(t))\frac{\partial}{\partial {\bar g}_{i}}
\right)
S(k; {\bar g}_{i}(t), {\bar L}(t))
&= 	0, \label{eq:3.9}
\end{align}
where the $\beta$-functions are defined by
\begin{align}
\beta_{g_{i}}({\bar g}_{i}(t))
&:= 	\frac{\partial {\bar g}_{i}(t)}{\partial t}\biggl|_{g_{i}, L_{0}}
\quad\text{with}\quad
{\bar g}_{i}(0)
= 	g_{i}, \label{eq:3.10}
\end{align}
which determines the running of $U(2)$ parameters.

In order to extract the $\beta$-functions we can differentiate the S-matrix in terms of $t$ explicitly:
\begin{align}
& 	\frac{\partial}{\partial t}S(k; {\bar g}_{i}(t), {\bar L}(t))\biggl|_{g_{i}, L_{0}} \nonumber\\
&= 	\begin{cases}
	\displaystyle
	-2\sum_{n=1}^{\infty}(-ik)^{n}\sum_{j=\pm}
	\left[
	n\bigl({\bar L}_{j}(t)\bigr)^{n-1}\frac{\partial {\bar L}_{j}(t)}{\partial t}{\bar P}_{j}(t)
	+ \bigl({\bar L}_{j}(t)\bigr)^{n}\frac{\partial {\bar P}_{j}(t)}{\partial t}
	\right]_{g_{i}, L_{0}}, \\
	\hfill\text{for~} |k\Bar{L}_{\pm}(t)|<1, \\[1ex]
	\displaystyle
	2\sum_{n=1}^{\infty}\left(\frac{i}{k}\right)^{n}\sum_{j=\pm}
	\left[
	-n\bigl({\bar L}_{j}(t)\bigr)^{-n-1}\frac{\partial {\bar L}_{j}(t)}{\partial t}{\bar P}_{j}(t)
	+ \bigl({\bar L}_{j}(t)\bigr)^{-n}\frac{\partial {\bar P}_{j}(t)}{\partial t}
	\right]_{g_{i}, L_{0}}, \\
	\hfill\text{for~} |k\Bar{L}_{\pm}(t)|>1,
	\end{cases}
\end{align}
where ${\bar L}_{\pm}(t) = {\bar L}(t)\cot({\bar \alpha}_{\pm}(t)/2)$, ${\bar P}_{\pm}(t) = (\1 \pm \Vec{\bar e}(t)\cdot\Vec{\sigma})/2$ and we have used the power series expansion
\begin{align}
S(k; \Bar{g}_{i}(t),\Bar{L}(t))
&= 	\begin{cases}
	\displaystyle-\1 - 2\sum_{n=1}^{\infty}\sum_{j=\pm}\bigl(-ik\Bar{L}_{j}(t)\bigr)^{n}\Bar{P}_{j}(t),
	& \text{for~} |k\Bar{L}_{\pm}(t)|<1, \\
	\displaystyle\1 + 2\sum_{n=1}^{\infty}\sum_{j=\pm}\left(\frac{i}{k\Bar{L}_{j}(t)}\right)^{n}\Bar{P}_{j}(t),
	& \text{for~} |k\Bar{L}_{\pm}(t)|>1.
	\end{cases} \label{eq:2.12}
\end{align}
In order to implement the requirement \eqref{eq:3.8}, the coefficient of $k^{n}$ must vanish for all $n$.
Furthermore, since ${\bar P}_{\pm}(t)$ are orthogonal to each other, it follows immediately that Eq.~\eqref{eq:3.9} will be satisfied if and only if the following conditions are fulfilled:
\begin{subequations}
\begin{align}
0
&= 	\frac{\partial {\bar L}_{\pm}(t)}{\partial t}\biggl|_{g_{i}, L_{0}}
= 	-{\bar L}_{\pm}(t)
	\left(1 + \frac{1}{\sin{\bar \alpha}_{\pm}(t)}\frac{\partial {\bar \alpha}_{\pm}(t)}{\partial t}\biggl|_{g_{i}, L_{0}}\right), \label{eq:3.13a}\\
0
&= 	\frac{\partial {\bar P}_{\pm}(t)}{\partial t}\biggl|_{g_{i}, L_{0}}
= 	\pm\frac{1}{2}\sum_{i=x,y,z}\frac{\partial {\bar e}_{i}(t)}{\partial t}\biggl|_{g_{i}, L_{0}}\sigma_{i}, \label{eq:3.13b}
\end{align}
\end{subequations}
from which we arrive at the \textit{exact} $\beta$-functions
\begin{subequations}
\begin{align}
\beta_{\alpha_{\pm}}({\bar \alpha}_{\pm}(t))
&= 	-\sin{\bar \alpha}_{\pm}(t), \label{eq:3.14a}\\
\beta_{e_{i}}({\bar e}_{i}(t))
&= 	0, \label{eq:3.14b}
\end{align}
\end{subequations}
where we have used the fact that each Pauli matrix is linearly independent.
From Eq.~\eqref{eq:3.14a} we see that ${\bar \alpha}_{\pm}(t) = 0$ and $\pi$ are a UV and an IR fixed point, respectively; see Fig.~\ref{fig:Beta}.

Let us next derive the running $U(2)$ parameters by using the exact $\beta$-functions.
To this end we integrate $\mathrm{d}t = \mathrm{d}\alpha_{\pm}/\beta_{\alpha_{\pm}}$ over the range $(0,t)$.
Noting that the initial value conditions ${\bar \alpha}_{\pm}(0) = \alpha_{\pm}$ we get
\begin{align}
\int_{0}^{t}\mathrm{d}{\tau}
&= 	-\int_{\alpha_{\pm}}^{{\bar \alpha}_{\pm}(t)}\!\!
	\frac{\mathrm{d}\alpha}{\sin\alpha}. \label{eq:3.15}
\end{align}
With the help of the integral formula $\int\!\frac{\mathrm{d}x}{\sin x} = \log|\tan\frac{x}{2}|$ we get
\begin{align}
t
= 	-\log\left|\frac{\tan({\bar \alpha}_{\pm}(t)/2)}{\tan(\alpha_{\pm}/2)}\right|, \label{eq:3.16}
\end{align}
from which we find $\bigl|\tan\frac{{\bar \alpha}_{\pm}(t)}{2}\bigr| = \bigl|\tan\frac{\alpha_{\pm}}{2}\bigr|\mathrm{e}^{-t}$, or, equivalently,
\begin{align}	
{\bar \alpha}_{\pm}(t)
&= 	2\arctan
	\left(
	\mathrm{e}^{-t}
	\tan\frac{\alpha_{\pm}}{2}
	\right). \label{eq:3.17}
\end{align}
Since the $\beta$-function for $e_{i}$ identically vanishes, its running becomes trivial:
\begin{align}
{\bar e}_{i}(t)
&= 	e_{i}. \label{eq:3.18}
\end{align}
All of these results are consistent with those obtained in the previous discussions.
\end{enumerate}

\subsection{Stability of fixed points} \label{sec:fixed_points}
\begin{figure}[t]
\centerline{\includegraphics{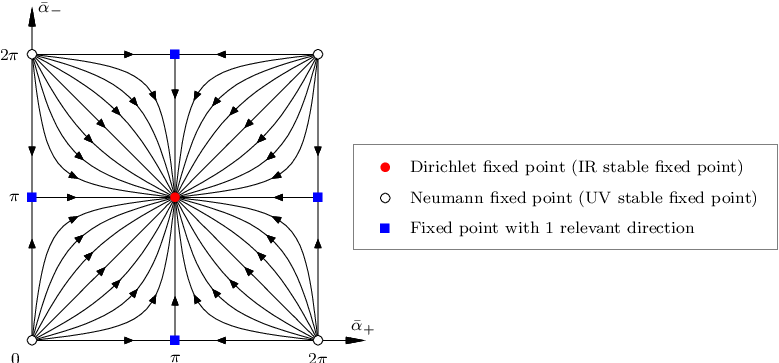}}
\caption{Exact RG flow of boundary conditions on the parameter space $(\alpha_{+}, \alpha_{-})$.
Arrows indicate the directions toward the infrared.}
\label{fig:RGflow}
\end{figure}

Let us next discuss the \textit{exact} RG flow in the $(\alpha_{+}, \alpha_{-})$-plane, which is a two dimensional torus $T^{2}$, by looking at the fixed points at which $\beta_{\alpha_{\pm}} = 0$ and analyze their stability.
To this end, let us first study the stability of a given fixed point.
As is well known, flow near a given fixed point follows from the linearized RG equations
\begin{align}
\left.
\frac{\partial}{\partial t}
\begin{pmatrix}
{\bar \alpha}_{+}(t) - \alpha_{+}^{\ast} \\
{\bar \alpha}_{-}(t) - \alpha_{-}^{\ast}
\end{pmatrix}
\right|_{\alpha_{j}, L_{0}}
&= 	M
	\begin{pmatrix}
	{\bar \alpha}_{+}(t) - \alpha_{+}^{\ast} \\
	{\bar \alpha}_{-}(t) - \alpha_{-}^{\ast}
	\end{pmatrix}
	+ O\bigl({\bar \alpha}_{\pm}(t) - \alpha_{\pm}^{\ast}\bigr)^{2}, \label{eq:3.19}
\end{align}
where $\alpha_{\pm}^{\ast} = 0$ or $\pi$, and $M$ is the \textit{stability matrix} given by
\begin{align}
M
&= 	\left.
	\begin{pmatrix}
	\frac{\partial \beta_{\alpha_{+}}}{\partial {\bar \alpha}_{+}} &
	\frac{\partial \beta_{\alpha_{+}}}{\partial {\bar \alpha}_{-}} \\[.5em]
	\frac{\partial \beta_{\alpha_{-}}}{\partial {\bar \alpha}_{+}} &
	\frac{\partial \beta_{\alpha_{-}}}{\partial {\bar \alpha}_{-}}
	\end{pmatrix}
	\right|_{{\bar \alpha}_{\pm} = \alpha_{\pm}^{\ast}}
= 	\begin{pmatrix}
	-\cos\alpha_{+}^{\ast} 	& 0 \\
	0 					& -\cos\alpha_{-}^{\ast}
	\end{pmatrix}. \label{eq:3.20}
\end{align}
The eigenvectors of $M$ with negative eigenvalues determine the relevant (IR unstable) directions at the given fixed point, and those with positive eigenvalues the irrelevant (IR stable) directions.
Since in our present case the stability matrix $M$ is already diagonal, we can easily check the stability of a given fixed point by looking at the explicit solution to the linearized equation
\begin{align}
\begin{pmatrix}
{\bar \alpha}_{+}(t) \\
{\bar \alpha}_{-}(t)
\end{pmatrix}
&= 	\begin{pmatrix}
	\alpha_{+}^{\ast} \\
	\alpha_{-}^{\ast}
	\end{pmatrix}
	+
	\begin{pmatrix}
	(\alpha_{+} - \alpha_{+}^{\ast})\exp[-(\cos\alpha_{+}^{\ast})t] \\
	(\alpha_{-} - \alpha_{-}^{\ast})\exp[-(\cos\alpha_{-}^{\ast})t]
	\end{pmatrix}
	+ O(\alpha_{\pm} - \alpha_{\pm}^{\ast})^{2}, \label{eq:3.21}
\end{align}
where we have imposed ${\bar \alpha}_{\pm}(0) = \alpha_{\pm}$.
Now it is easy to see the relevancy or irrelevancy of a given fixed point $(\alpha_{+}^{\ast}, \alpha_{-}^{\ast})$ (see for textbook exposition Ref.~\citen{Cardy:1996xt}):
\begin{itemize}
\item If $\alpha_{j}^{\ast} = 0$ ($j=+$ or $-$), $\alpha_{j}$-direction is relevant: as lower the energy scale $t \to -\infty$, the running eigenphase ${\bar \alpha}_{j}(t)$ moves away from its fixed point value $\alpha_{j}^{\ast}$ exponentially.

\item If $\alpha_{j}^{\ast} = \pi$, $\alpha_{j}$-direction is irrelevant: as lower the energy scale $t \to -\infty$, the running eigenphase ${\bar \alpha}_{j}(t)$ moves towards its fixed point value $\alpha_{j}^{\ast}$ exponentially.
\end{itemize}
We may distinguish $2^{2} = 4$ fixed points by the number of relevant directions.
As depicted in Fig.~\ref{fig:RGflow}, there exist the following three distinct types of fixed points:\footnote{A renormalization group analysis of $U(2)$ family of boundary conditions
has been given in Refs.~\citen{Asorey:2007rt,Asorey:2007kw} from a field theoretical point of view, but our results are different from those advocated in the literature\cite{Asorey:2007rt,Asorey:2007kw}, where IR fixed point corresponds to Neumann boundary condition and UV fixed point Dirichlet boundary condition.}

\begin{enumerate}
\item \textit{$(0, 0)$-fixed point (UV stable Neumann fixed point)}.\\
There is a ultraviolet stable fixed point at $(\alpha_{+}, \alpha_{-}) = (0, 0)$.
At this point the unitary matrix becomes the identity matrix, $U = P_{+} + P_{-} = \1$.
Thus, by referring to the equation \eqref{eq:1.7}, we see that this fixed point corresponds to the Neumann-Neumann boundary conditions at the origin:
\begin{align}
\psi^{\prime}(0_{-})
&= 	0
= 	\psi^{\prime}(0_{+}). \label{eq:3.22}
\end{align}

\item \textit{$(\pi, \pi)$-fixed point (IR stable Dirichlet fixed point)}.\\
There is an infrared stable fixed point at $(\alpha_{+}, \alpha_{-}) = (\pi, \pi)$.
At this point the unitary matrix is $U = -P_{+} - P_{-} = -\1$.
Thus, from Eq.~\eqref{eq:1.7} we see that this fixed point corresponds to the Dirichlet-Dirichlet boundary conditions:
\begin{align}
\psi(0_{-})
&= 	0
= 	\psi(0_{+}). \label{eq:3.23}
\end{align}

\item \textit{$(0, \pi)$- and $(\pi, 0)$-fixed points}.\\
There are other fixed points at $(\alpha_{+}, \alpha_{-}) = (0, \pi)$ and $(\pi, 0)$, at which the unitary matrix becomes $U = \pm P_{+} \mp P_{-} = \pm \Vec{e}\cdot\Vec{\sigma}$.
Since these two fixed points are related by the exchange $\Vec{e} \leftrightarrow -\Vec{e}$, in the following we will concentrate on the case $(\alpha_{+}, \alpha_{-}) = (0, \pi)$.
This fixed point is IR stable in the $\alpha_{-}$-direction and unstable only in the $\alpha_{+}$-direction; see Fig.~\ref{fig:RGflow}.
Referring to the equations \eqref{eq:1.11a} and \eqref{eq:1.11b}, we see that this fixed point corresponds to the boundary conditions $P_{-}\Vec{\Psi}(0_{+}) = \Vec{0}$ and $P_{+}\Vec{\Psi}^{\prime}(0_{+}) = \Vec{0}$.
With the parameterization
\begin{align}
\Vec{e}
&= 	(\sin\theta\cos\varphi, \sin\theta\sin\varphi, \cos\theta)^{T}, \quad
0\leq\theta\leq\pi, \quad
0\leq\varphi<2\pi, \label{eq:3.24}
\end{align}
the projection operators become
\begin{subequations}
\begin{align}
P_{+}
&= 	\frac{1}{2}
	\begin{pmatrix}
	1+\cos\theta 				& \mathrm{e}^{-i\varphi}\sin\theta \\
	\mathrm{e}^{i\varphi}\sin\theta 	& 1-\cos\theta
	\end{pmatrix}
= 	\begin{pmatrix}
	\cos^{2}\frac{\theta}{2} 								& \mathrm{e}^{-i\varphi}\sin\frac{\theta}{2}\cos\frac{\theta}{2} \\[.1em]
	\mathrm{e}^{i\varphi}\sin\frac{\theta}{2}\cos\frac{\theta}{2} 	& \sin^{2}\frac{\theta}{2}
	\end{pmatrix}, \label{eq:3.25a}\\
P_{-}
&= 	\frac{1}{2}
	\begin{pmatrix}
	1-\cos\theta 					& -\mathrm{e}^{-i\varphi}\sin\theta \\
	-\mathrm{e}^{i\varphi}\sin\theta 	& 1+\cos\theta
	\end{pmatrix}
= 	\begin{pmatrix}
	\sin^{2}\frac{\theta}{2} 							& -\mathrm{e}^{-i\varphi}\sin\frac{\theta}{2}\cos\frac{\theta}{2} \\[.1em]
	-\mathrm{e}^{i\varphi}\sin\frac{\theta}{2}\cos\frac{\theta}{2} 	& \cos^{2}\frac{\theta}{2}
	\end{pmatrix}. \label{eq:3.25b}
\end{align}
\end{subequations}
Thus, with this parameterization, the boundary conditions $P_{-}\Vec{\Psi}(0_{+}) = \Vec{0}$ and $P_{+}\Vec{\Psi}^{\prime}(0_{+}) = \Vec{0}$ are cast into the following forms:
\begin{subequations}
\begin{align}
\psi(0_{-})
&= 	\mathrm{e}^{i\varphi}\tan\tfrac{\theta}{2}\psi(0_{+}), \label{eq:3.26a}\\
\psi^{\prime}(0_{-})
&= 	\mathrm{e}^{i\varphi}\cot\tfrac{\theta}{2}\psi^{\prime}(0_{+}), \label{eq:3.26b}
\end{align}
\end{subequations}
which are the scale-independent boundary conditions discussed in Refs.~\citen{Albeverio:1998,Fulop:1999pf,Cheon:2000tq} and \citen{Fulop:2003}.

As we have seen in \S \ref{sec:beta-function}, $\Vec{e}$ is the exactly marginal parameter such that $\theta$ and $\varphi$ are just determined by its initial values, or the microscopic cutoff theory.
Although these marginal parameters do not flow against the RG, they can be restricted by symmetry.
Since in a philosophy of RG, low energy physics will be well captured by symmetry and relevant as well as marginal parameters, it is very important how these marginal parameters are restricted by symmetry.
Before closing this section let us take a look at the restrictions on $\theta$ and $\varphi$ by the symmetry requirements.
\begin{itemize}
\item \textit{Parity invariant case}.\\
Let us first study the parity invariant case of $(0, \pi)$-fixed point.
It is known that the parity invariant subfamily of point interactions is specified by the constraint $U = \sigma_{1}U\sigma_{1}$. \cite{Albeverio:1998,Cheon:2000tq,Fulop:2003}
The solution to this requirement is given by $\Vec{e} = (\pm 1, 0, 0)^{T}$, or $(\theta, \varphi) = (\frac{\pi}{2}, 0), (\frac{\pi}{2}, \pi)$ in \eqref{eq:3.24}.
Thus, for $\Vec{e} = (+1, 0, 0)^{T}$, the boundary condition for $(0, \pi)$-fixed point reduces to
\begin{subequations}
\begin{align}
\psi(0_{-})
&= 	\psi(0_{+}), \label{eq:3.27a}\\
\psi^{\prime}(0_{-})
&= 	\psi^{\prime}(0_{+}), \label{eq:3.27b}
\end{align}
\end{subequations}
which is the perfectly connected boundary condition that describes the "free theory".
For $\Vec{e} = (-1, 0, 0)^{T}$, on the other hand, the boundary conditions \eqref{eq:3.26a} and \eqref{eq:3.26b} reduce to the following:
\begin{subequations}
\begin{align}
\psi(0_{-})
&= 	-\psi(0_{+}), \label{eq:3.28a}\\
\psi^{\prime}(0_{-})
&= 	-\psi^{\prime}(0_{+}). \label{eq:3.28b}
\end{align}
\end{subequations}

\item \textit{Time-reversal invariant case}.\\
Let us next consider the time-reversal invariant case of $(0, \pi)$-fixed point.
The time-reversal invariant subfamily of point interactions is specified by the constraint $U = U^{T}$. \cite{Albeverio:1998,Cheon:2000tq,Fulop:2003}
The solution to this requirement is given by $\Vec{e} = (e_{x}, 0, e_{z})^{T}$, or $\varphi = 0, \pi$ in \eqref{eq:3.24}.
The boundary conditions \eqref{eq:3.26a} and \eqref{eq:3.26b} thus become
\begin{subequations}
\begin{align}
\psi(0_{-})
&= 	\pm\tan\tfrac{\theta}{2}\psi(0_{+}), \label{eq:3.29a}\\
\psi^{\prime}(0_{-})
&= 	\pm\cot\tfrac{\theta}{2}\psi^{\prime}(0_{+}), \label{eq:3.29b}
\end{align}
\end{subequations}
where `$+$'-sign for $\varphi = 0$ and `$-$'-sign for $\varphi = \pi$.
This is the boundary condition realized by the system of scale-independent $\delta^{\prime}$ interaction described by the Hamiltonian $H = -\frac{\mathrm{d}^{2}}{\mathrm{d}x^{2}} + 2\frac{1 \mp \tan(\theta/2)}{1 \pm \tan(\theta/2)}\delta^{\prime}(x)$. \cite{Griffiths:1993,Kurasov:1994}
We note that when $\theta = 0$ this boundary condition boils down to the Dirichlet-Neumann boundary condition $\psi(0_{-}) = 0 = \psi^{\prime}(0_{+})$, while when $\theta = \pi$ it becomes the Neumann-Dirichlet boundary condition $\psi^{\prime}(0_{-}) = 0 = \psi(0_{+})$.

\item \textit{$\mathcal{PT}$-symmetric case}.\\
Let us finally consider the $\mathcal{PT}$-symmetric case of $(0, \pi)$-fixed point, where $\mathcal{PT}$ is the composite operation of parity and time-reversal.
It is known that $\mathcal{PT}$-symmetric subfamily of point interactions is specified by the constraint $U = \sigma_{1}U^{T}\sigma_{1}$. \cite{Albeverio:1998,Cheon:2000tq,Fulop:2003}
The solution to this requirement is given by $\Vec{e} = (e_{x}, e_{y}, 0)^{T}$, or $\theta = \frac{\pi}{2}$ in \eqref{eq:3.24}.
Then the boundary conditions \eqref{eq:3.26a} and \eqref{eq:3.26b} reduce to the well-known twisted boundary conditions
\begin{subequations}
\begin{align}
\psi(0_{-})
&= 	\mathrm{e}^{i\varphi}\psi(0_{+}), \label{eq:3.30a}\\
\psi^{\prime}(0_{-})
&= 	\mathrm{e}^{i\varphi}\psi^{\prime}(0_{+}). \label{eq:3.30b}
\end{align}
\end{subequations}
\end{itemize}
\end{enumerate}

\section{Duality} \label{sec:duality}
\begin{figure}[t]
\begin{center}
\begin{tabular}{ccc}
\parbox{4cm}{\includegraphics{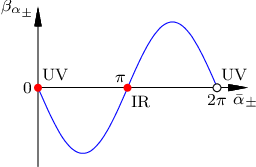}}
&
$\xrightarrow[\text{UV} \leftrightarrow \text{IR}]{\text{duality transformation}}$
&
\parbox{4cm}{\includegraphics{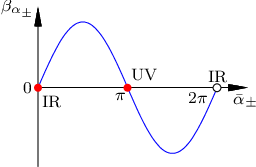}}
\end{tabular}
\end{center}
\caption{$\beta$-function flip under the duality transformation \eqref{eq:4.3}. UV and IR fixed points are interchanged.}
\label{fig:dual}
\end{figure}
It is well-known that there exists a remarkable identity in the S-matrix which signifies an equivalence between UV ($k \to \infty$) and IR ($k \to 0$) regimes of two distinct theories provided by two different boundary conditions.
Such an identity has been studied in the literature under the name of duality, especially in the context of quantum graph (see for example Ref.~\citen{Kostrykin:1999}).
In this section we will briefly discuss yet another aspect of duality from the viewpoint of RG.

To this end we first note that the eigenvalue of the S-matrix satisfies the following elementary identity:
\begin{align}
\frac{ikL_{0}\cot\frac{\alpha_{j}}{2} - 1}{ikL_{0}\cot\frac{\alpha_{j}}{2} + 1}
&= 	- \frac{i(kL_{0})^{-1}\cot\frac{\alpha_{j}\pm\pi}{2} - 1}
	{i(kL_{0})^{-1}\cot\frac{\alpha_{j}\pm\pi}{2} + 1} \label{eq:4.1},
\end{align}
from which we find
\begin{align}
S(kL_{0}; \alpha_{j}, e_{j})
&= 	- S((kL_{0})^{-1}; \alpha_{j}\pm\pi, e_{j}). \label{eq:4.2}
\end{align}
Eq.~\eqref{eq:4.1} indicates that the high-energy regime of the system characterized by the parameters $\{\alpha_{j}, e_{j}\}$ is equivalent, or \textit{dual} to the (opposite sign of) low-energy regime of that characterized by $\{\alpha_{j} \pm \pi, e_{j}\}$.
Since $L_{\pm}$ has different sign in these two different systems, we immediately see that a normalizable bound state in one system is dual to a non-normalizable antibound state in the other.

Since the duality \eqref{eq:4.2} connects the high energy and low energy regimes of two different theories, it would be reasonable to expect that there would exist a duality between UV and IR fixed points.
Indeed, if we consider the following transformation:
\begin{align}
D:
{\bar \alpha}_{j}
\mapsto
\begin{cases}
{\bar \alpha}_{j} + \pi, 	& \text{for~} 0\leq\alpha_{j}<\pi, \\
{\bar \alpha}_{j} - \pi, 	& \text{for~} \pi\leq\alpha_{j}<2\pi,
\end{cases} \label{eq:4.3}
\end{align}
which acts on the S-matrix as
\begin{align}
S(kL_{0}; \alpha_{j}, e_{j})
&\stackrel{D}{\mapsto}
	- S((kL_{0})^{-1}; \alpha_{j}, e_{j}), \label{eq:4.4}
\end{align}
the $\beta$-functions flip its sign under $D$ and hence the UV and IR fixed points are interchanged (see Fig.~\ref{fig:dual}).
In this sense $(\pi, \pi)$-fixed point (Dirichlet fixed point) is dual to $(0, 0)$-fixed point (Neumann fixed point) and $(0, \pi)$-fixed point is dual to $(\pi, 0)$-fixed point.

In summary, the duality consists of the followings:
\begin{itemize}
\item \textit{High energy/low energy scattering duality}.\\
Scattering process in high energy regime ($|kL_{\pm}|>1$) of the system characterized by the parameters $\{\alpha_{j}, e_{j}\}$ is dual to low energy regime ($|kL_{\pm}|<1$) of that characterized by $\{\alpha_{j} \pm \pi, e_{j}\}$.

\item \textit{Bound state/antibound state duality}.\\
A normalizable bound state characterized by a simple pole $kL_{0} = i\tan\frac{\alpha_{\pm}}{2}$ lying on the positive imaginary $k$-axis is dual to a non-normalizable antibound state characterized by a simple pole $kL_{0} = -i\cot\frac{\alpha_{\pm}}{2}$ lying on the imaginary $k$-axis (where we have assumed that $L_{0}>0$ and $0<\alpha_{\pm}<\pi$).

\end{itemize}

\section{Conclusions and discussions} \label{sec:conclusions}
In this paper we argued that boundary condition may flow with energy scale if we regard it as a low-energy effective description for some underlying short-ranged interaction.
As the simplest example we studied the RG flow of $U(2)$ family of boundary conditions in the framework of one-particle quantum mechanics.
We required that physical quantities, i.e. the S-matrix and bound state energies should not depend on the choice of the arbitrary length scale $L_{0}$, which is inevitably introduced into the boundary condition on account of dimensional analysis.
With this requirement we arrived at the exact $\beta$-functions by differentiating the S-matrix or bound state energies in terms of RG time $t$ explicitly.
It should be noted here that since $L_{0}$ always appears in the combination $L_{\pm} = L_{0}\cot\frac{\alpha_{\pm}}{2}$, our RG equations \eqref{eq:3.2a} and \eqref{eq:3.2b} or its differential form \eqref{eq:3.9} are just equivalent to the following simple equation:
\begin{align}
L_{0}\cot\frac{\alpha_{\pm}}{2}
&= 	\Bar{L}(t)\cot\frac{\Bar{\alpha}_{\pm}(t)}{2}, \label{eq:5.1}
\end{align}
where $\Bar{L}(t)$ is given in \eqref{eq:3.1}.
By solving the equation \eqref{eq:5.1} we showed that there are three distinct fixed points, where the Weyl rescaling invariant boundary conditions \cite{Albeverio:1998,Fulop:1999pf,Cheon:2000tq,Fulop:2003} are realized.
This result strongly suggests that in one spatial dimension there exist three types of universality classes of short-ranged interaction:
If UV theory lies on the critical point $(\alpha_{+}, \alpha_{-}) = (0, 0)$ it remains on the Neumann fixed point.
If UV theory lies on the critical lines $\alpha_{+} = 0$ ($\alpha_{-} = 0$), it flows into $(0, \pi)$-fixed point ($(\pi, 0)$-fixed point).
All other short-ranged interactions flow into the Dirichlet fixed point in the long-wavelength limit, which implies that without fine-tuning most of localized potentials will be effectively described by an infinitely deep Dirichlet wall in the low-energy limit.\footnote{It is known that fine-tuning is necessary to obtain anything but the Dirichlet wall in the context of one-particle quantum mechanics on a half-line with a piece-wise flat potential and/or Morse potential.\cite{Fulop:2002,Belchev:2010}
These results are quite in agreement with our RG analysis.}

We are left with a number of questions, however.
One of main questions is how our RG flow will be modified when bulk interactions are added.
Let us close this paper by looking at possible modifications of running boundary condition in conformal mechanics and in quantum graph.

\subsection{Conformal mechanics}
In this paper we focused on systems of short-ranged interaction and its long-wavelength limit, however, it is also interesting to investigate systems governed by long-ranged interaction in the bulk, whose long-wavelength limit cannot be reduced to a point interaction.
One of typical examples for such systems is the one-dimensional conformal mechanics \cite{deAlfaro:1976je} described by the bulk Hamiltonian
\begin{align}
H_{\text{bulk}}
&= 	- \frac{\mathrm{d}^{2}}{\mathrm{d}r^{2}} + \frac{g}{r^{2}}, \quad
0<r<\infty, \label{eq:5.2}
\end{align}
where $g$ is a dimensionless coupling constant.
As in the case of free Hamiltonian, \eqref{eq:5.2} is characterized by lack of any scale parameter.
However, boundary condition at the origin may introduce a new length scale $L_{0}$.
Indeed, for the coupling constant lying on the range $-1/4<g<3/4$,\footnote{The upper bound $g=3/4$ is due to the reason that one of the two independent zero-energy solutions $\psi(r) \propto r^{\nu_{-}}$ becomes non-square-integrable at the origin when $g \geq 3/4$. Hence in the region $3/4 \leq g < \infty$ there is no chance for the self-adjoint extension of \eqref{eq:5.2}.} it can be shown that the self-adjoint extension of $H_{\text{bulk}}$ admits the following $U(1)$ family of boundary conditions (see Appendix \ref{appendix}):
\begin{align}
\left[
\frac{\psi(r)}{r^{\nu_{-}}}
- iL_{0}^{1-2\nu_{-}}r^{2\nu_{-}}\frac{\mathrm{d}}{\mathrm{d}r}\frac{\psi(r)}{r^{\nu_{-}}}
\right]_{r=0}
&= 	\mathrm{e}^{i\alpha}
	\left[
	\frac{\psi(r)}{r^{\nu_{-}}}
	+ iL_{0}^{1-2\nu_{-}}r^{2\nu_{-}}\frac{\mathrm{d}}{\mathrm{d}r}\frac{\psi(r)}{r^{\nu_{-}}}
	\right]_{r=0}, \label{eq:5.3}
\end{align}
or, equivalently,
\begin{align}
\left[
\frac{\psi(r)}{r^{\nu_{-}}}
+ L_{0}^{\nu_{+} - \nu_{-}}\left(\cot\frac{\alpha}{2}\right)r^{2\nu_{-}}\frac{\mathrm{d}}{\mathrm{d}r}\frac{\psi(r)}{r^{\nu_{-}}}
\right]_{r=0}
&= 	0, \label{eq:5.4}
\end{align}
where $0 \leq \alpha < 2\pi$ and $\nu_{\pm} := 1/2 \pm \sqrt{g + 1/4}$.
Just as in the case of free Hamiltonian, the scale parameter $L_{0}$ appears with the combination $L_{0}^{\nu_{+} - \nu_{-}}\cot\frac{\alpha}{2}$.
Thus, on the analogy of \eqref{eq:5.1}, the RG equation must be of the form\footnote{The RG equation \eqref{eq:5.5} has been advocated before by Amelino-Camelia and Bak \cite{AmelinoCamelia:1994we} (see also Ref.~\citen{AmelinoCamelia:1997hh}) in the context of $(2+1)$-dimensional Chern-Simons field theory.}
\begin{align}
L_{0}^{\nu_{+} - \nu_{-}}\cot\frac{\alpha}{2}
&= 	\bigl[\Bar{L}(t)\bigr]^{\nu_{+} - \nu_{-}}\cot\frac{\Bar{\alpha}(t)}{2}. \label{eq:5.5}
\end{align}
The RG equation \eqref{eq:5.5} is easily solved with the result $\Bar{\alpha}(t) = 2\arctan(\mathrm{e}^{-(\nu_{+} - \nu_{-})t}\tan\frac{\alpha}{2})$ and the differentiation of \eqref{eq:5.5} with respect to $t$ leads to the following \textit{exact} $\beta$-function:\footnote{RG approaches to conformal mechanics based on the regularization of inverse-squared potential are found in Refs.~\citen{Gupta:1993id,Beane:2000wh,Barford:2002je,Bawin:2003dm,Camblong:2003mb,Barford:2004fz,Kaplan:2009kr,Moroz:2009nm,Moroz:2009kv}}
\begin{align}
\beta_{\alpha}(\Bar{\alpha}(t))
&:= 	\frac{\partial\Bar{\alpha}(t)}{\partial t}\biggr|_{\alpha, g, L_{0}}
= 	- (\nu_{+} - \nu_{-})\sin\Bar{\alpha}(t). \label{eq:5.6}
\end{align}
It should be emphasized that the $\beta$-function \eqref{eq:5.6} is the same as that for free Hamiltonian except for the overall factor $(\nu_{+} - \nu_{-}) = 2\sqrt{g + 1/4}$, which describes the effect of bulk interaction.
Consequently, the RG flow of boundary condition is almost the same to the free Hamiltonian case and thus there exist two fixed points $\alpha^{\ast} = 0$ and $\pi$, the former is UV stable and the latter IR stable.
At the UV fixed point $\alpha^{\ast} = 0$ the Neumann type boundary condition $r^{2\nu_{-}}(\mathrm{d}/\mathrm{d}r)(\psi(r)/r^{\nu_{-}})|_{r=0} = 0$ is realized.
At the IR fixed point $\alpha^{\ast} = \pi$, on the other hand, the Dirichlet type boundary condition $\psi(r)/r^{\nu_{-}}|_{r=0} = 0$ is realized.
Impacts of these two fixed points are already familiar in the context of AdS/CFT:\cite{Klebanov:1999tb}
A massive scalar field theory on AdS$_{d+1}$ with the Euclidean conformal metric $\mathrm{d}s^{2} = (R/z)^{2}(\mathrm{d}z^{2} + \sum_{i=1}^{d}\mathrm{d}x_{i}^{2})$ (where $R$ is the AdS curvature scale) can produce CFT operators  with two distinct scaling dimensions if the bulk scalar mass $m$ lies on the range $-1/4 < (d^{2} - 1)/4 + (mR)^{2} < 3/4$,\cite{Klebanov:1999tb} which is just the consequence of the presence of UV and IR fixed points in the region $-1/4 < g < 3/4$ of one-dimensional conformal mechanics.

Conformal mechanics is merely one of examples, however, the above results may imply that regardless of the presence of bulk interaction the RG flow of boundary conditions would be almost the same to the system of point interactions:
In the low-energy limit the most stable boundary condition in one spatial dimension would be the Dirichlet (type) boundary condition.

\subsection{Quantum graph}
\begin{wrapfigure}{r}{6cm}
\centerline{\includegraphics{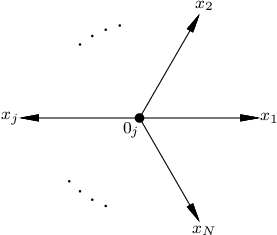}}
\caption{Star graph with $N$ edges.}
\label{fig:6}
\end{wrapfigure}
It is easy to generalize our analysis to quantum graph, which is an idealized one-dimensional system whose geometry is a graph, i.e. a set of finite and/or semi-infinite lines that are connected at some vertices by certain relations (see for example Ref.~\citen{Kostrykin:1999}).
The simplest quantum graph is a star graph, which consists of several half-lines (usually referred to as edges) joining at a single vertex; see Fig.~\ref{fig:6}.
In quantum mechanics for a free particle on a star graph with $N$ edges, a similar analysis to that presented in \S \ref{sec:intro} shows that the vertex consistent with the provability current conservation $\sum_{j=1}^{N}j(0_{j}) = 0$ is described by the following $U(N)$ family of boundary conditions ($U \in U(N)$)
\begin{align}
\Vec{\Psi}(0_{1},\cdots,0_{N}) - iL_{0}\Vec{\Psi}^{\prime}(0_{1},\cdots,0_{N})
&= 	U\bigl[\Vec{\Psi}(0_{1},\cdots,0_{N}) + iL_{0}\Vec{\Psi}^{\prime}(0_{1},\cdots,0_{N})\bigr], \label{eq:5.7}
\end{align}
where the local provability current density on the $j$th edge is given by $j(x_{j}) = -i[\psi^{\prime\ast}(x_{j})\psi(x_{j}) - \psi^{\ast}(x_{j})\psi^{\prime}(x_{j})]$, $\Vec{\Psi}$ and $\Vec{\Psi}^{\prime}$ are $N$-component column vectors defined by
$\Vec{\Psi}(x_{1}, \cdots, x_{N})
:= 	\bigl(
	\psi(x_{1}), \cdots, \psi(x_{N})
	\bigr)^{T}$ and
$\Vec{\Psi}^{\prime}(x_{1}, \cdots, x_{N})
:= 	\bigl(
	\psi^{\prime}(x_{1}), \cdots, \psi^{\prime}(x_{N})
	\bigr)^{T}$.
$x_{j}$ ($0_{j} \leq x_{j}<\infty$; $j=1,\cdots,N$) is the coordinate on the $j$th edge and $\psi(x_{j})$ is the wavefunction on the $j$th edge.
The model analyzed in the previous sections is a special case of the star graph with $N=2$.

\begin{figure}[t]
\centerline{\includegraphics{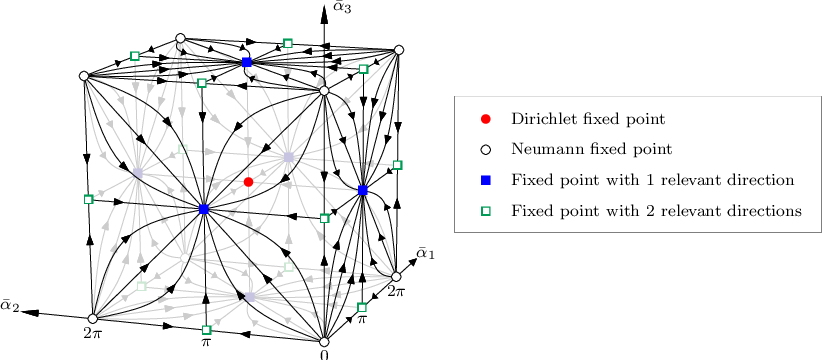}}
\caption{Exact RG flow of boundary conditions at the vertex of star graph with $N=3$ edges (so-called Y-junction) on the parameter space $(\alpha_{1}, \alpha_{2}, \alpha_{3})$. Arrows indicate the directions toward the infrared. There exist $2^{3} = 8$ distinct fixed points.}
\label{fig:7}
\end{figure}

By making use of the spectral decomposition $U = \sum_{j=1}^{N}\mathrm{e}^{i\alpha_{j}}P_{j}$, the boundary condition  \eqref{eq:5.7} boils down to the following $N$ independent conditions:
\begin{align}
P_{j}\left[\Vec{\Psi}(0_{1},\cdots,0_{N})
+ L_{0}\left(\cot\frac{\alpha_{j}}{2}\right)\Vec{\Psi}^{\prime}(0_{1},\cdots,0_{N})\right]
&= 	\Vec{0}, \quad
j=1,\cdots,N, \label{eq:5.8}
\end{align}
which show that the arbitrary scale parameter $L_{0}$ again appears only with the combination $L_{0}\cot(\alpha_{j}/2)$.
The RG flow of the vertex is thus determined by the running eigenphase $\Bar{\alpha}_{j}(t)$ subject to the RG equation $L_{0}\cot(\alpha_{j}/2) = \Bar{L}(t)\cot(\Bar{\alpha}_{j}(t)/2)$, from which we find the following \textit{exact} $\beta$-function:
\begin{align}
\beta_{\alpha_{j}}(\Bar{\alpha}_{j}(t))
&:= 	\frac{\partial\Bar{\alpha}_{j}(t)}{\partial t}\biggr|_{\alpha_{j}, P_{j}, L_{0}}
= 	- \sin\Bar{\alpha}_{j}(t), \quad
j=1,\cdots,N, \label{eq:5.9}
\end{align}
which has two zero points $\alpha_{j}^{\ast} = 0$ and $\pi$.
Thus, the resultant RG flow admits $2^{N}$ distinct fixed points, among which the most IR (UV) stable fixed point is again the Dirichlet (Neumann) fixed point described by the boundary condition $\psi(0_{1}) = \cdots = \psi(0_{N}) = 0$ ($\psi^{\prime}(0_{1}) = \cdots = \psi^{\prime}(0_{N}) = 0$).
The other fixed points possess the relevant directions.
We note that the number of fixed points with $n$ relevant directions is given by the combination $\binom{N}{n} = \frac{N!}{n!(N-n)!}$.
(The total number of fixed points is thus given by $\sum_{n=0}^{N}\binom{N}{n} = (1 + 1)^{N} = 2^{N}$, as it should.)
As an illustrative example, we depict the exact RG flow of boundary conditions for the case of $N=3$ in Fig.~\ref{fig:7}.

It should be mentioned here that our RG flow of boundary conditions depicted in Fig.~\ref{fig:7} just coincides with the RG flow of a junction of three quantum wires given in the context of Tomonaga-Luttinger (TL) liquid on star graph\cite{Bellazzini:2008fu} (see also Ref.~\citen{Bellazzini:2009nk}).
There, the scaling dimensions of fermion two-point functions are, in roughly speaking, given by $(\text{universal bulk dimension}) + (\text{boundary dimension})$, the former includes only the bulk coupling constants (and hence is universal) and the latter the parameters specifying boundary condition at the vertex.
Furthermore, the boundary dimensions of the two-point functions are given by $(\text{bulk coupling constants}) \times (\text{eigenvalues of }U)$, where $U \in U(N)$ is the same unitary matrix as ours (where in Ref.~\citen{Bellazzini:2008fu} the time-reversal invariance is assumed).
Thus, up to an overall factor, the RG flow of a junction in TL liquid follows from the same flow as the quantum mechanical one depicted in Fig.~\ref{fig:7}.
It is very similar to the case of conformal mechanics that the bulk interaction effects only appear in the overall factor.
However, we do not know whether there is a deeper reason behind this similarity or not.

\section*{Acknowledgment}
We would like to thank T. Cheon, M. Mintchev, S. Odake and I. Tsutsui for useful discussions and comments.
SO is supported in part by JSPS Research Fellowships for Young Scientists and the Excellent Young Researchers Overseas Visit Program conducted by JSPS.
MS is supported in part by the Grant-in-Aid for Scientific Research (No.1854075) by the Japanese Ministry of Education, Science, Sports and Culture.

\appendix
\section{Boundary condition of conformal mechanics} \label{appendix}
In this appendix we show that when $g>-1/4$ the self-adjoint extension of the bulk Hamiltonian \eqref{eq:5.2} admits the $U(1)$ family of boundary conditions given in \eqref{eq:5.4}.

To this end, let us consider the following time-independent Schr\"odinger equation:
\begin{align}
H_{\text{bulk}}\psi(r)
&= 	E\psi(r), \quad
0<r<\infty, \label{eq:1}
\end{align}
with the bulk Hamiltonian $H_{\text{bulk}} = -\mathrm{d}^{2}/\mathrm{d}r^{2} + g/r^{2}$.
The general solution to the equation \eqref{eq:1} obeys the following short-distance ($r \to 0$) behavior
\begin{align}
\psi(r)
&= 	r^{\nu_{+}}\bigl[\text{const.} + O(r^{2})\bigr]
	+ r^{\nu_{-}}\bigl[\text{const.} + O(r^{2})\bigr], \label{eq:3}
\end{align}
where $\nu_{\pm} = 1/2 \pm \sqrt{g+1/4}$.
Notice that $r^{\nu_{+}}$ and $r^{\nu_{-}}$ are two independent zero-energy solutions to the equation \eqref{eq:1}.\footnote{When $g=-1/4$ the two independent zero-energy solutions are $r^{1/2}$ and $r^{1/2}\log (r/L_{0})$, where $L_{0}$ is an arbitrary length parameter which has to be introduced in order to guarantee the argument of the logarithm to be dimensionless.}
A delicate problem arises when one specifies the boundary condition for $\psi$ because $r^{\nu_{-}}$ diverges at the origin and hence the wavefunction $\psi$ itself has no definite boundary value in general.
In order to resolve this problem, we first note that the bulk Hamiltonian can be written as follows:\begin{align}
H_{\text{bulk}}
&= 	\left(-\frac{\mathrm{d}}{\mathrm{d}r} - \frac{\nu}{r}\right)
	\left(\frac{\mathrm{d}}{\mathrm{d}r} - \frac{\nu}{r}\right)
= 	-\frac{1}{r^{\nu}}\frac{\mathrm{d}}{\mathrm{d}r}r^{2\nu}
	\frac{\mathrm{d}}{\mathrm{d}r}\frac{1}{r^{\nu}}, \label{eq:4}
\end{align}
where $\nu$ is either $\nu_{+}$ or $\nu_{-}$.
For the following discussions, however, we choose $\nu = \nu_{-}$ in order to divide the singularity due to the zero-energy solution $r^{\nu_{-}}$.

Next study the boundary condition for the wavefunctions, which has to be consistent with the self-adjointness of the bulk Hamiltonian \eqref{eq:4}.
To this end we first compute the following quantity:
\begin{align}
\Omega(\varphi, \psi)
&:=	\langle \varphi|H_{\text{bulk}}\psi\rangle
	- \langle H_{\text{bulk}}\varphi|\psi\rangle \nonumber\\
&= 	\int_{0}^{\infty}\!\!\mathrm{d}r
	\left[
	\varphi^{\ast}(r)
	\left(-\frac{1}{r^{\nu}}\frac{\mathrm{d}}{\mathrm{d}r}r^{2\nu}
	\frac{\mathrm{d}}{\mathrm{d}r}\frac{\psi(r)}{r^{\nu}}\right)
	- \left(-\frac{1}{r^{\nu}}\frac{\mathrm{d}}{\mathrm{d}r}r^{2\nu}
	\frac{\mathrm{d}}{\mathrm{d}r}\frac{\varphi(r)}{r^{\nu}}\right)^{\ast}
	\psi(r)
	\right] \nonumber\\
&= 	\left[
	- \left(\frac{\varphi(r)}{r^{\nu}}\right)^{\ast}r^{2\nu}
	\frac{\mathrm{d}}{\mathrm{d}r}\frac{\psi(r)}{r^{\nu}}
	+ \left(r^{2\nu}\frac{\mathrm{d}}{\mathrm{d}r}\frac{\varphi(r)}{r^{\nu}}\right)^{\ast}
	\frac{\psi(r)}{r^{\nu}}
	\right]_{r=0}, \label{eq:5}
\end{align}
where in the third equality we have integrated by parts and further assumed that $(1/r^{\nu})^{\ast} = 1/r^{\nu}$, that is, $g>-1/4$.
The self-adjointness of $H_{\text{bulk}}$ requires $\Omega(\varphi, \psi) = 0$ for any $\varphi$ and $\psi$.
We rewrite the requirement $\Omega(\varphi, \psi) = 0$ into the following way:
\begin{align}
& 	\left[
	\frac{\varphi(r)}{r^{\nu}} - iL_{0}^{1 - 2\nu}r^{2\nu}
	\frac{\mathrm{d}}{\mathrm{d}r}\frac{\varphi(r)}{r^{\nu}}
	\right]_{r=0}^{\ast}
	\left[
	\frac{\psi(r)}{r^{\nu}} - iL_{0}^{1 - 2\nu}r^{2\nu}
	\frac{\mathrm{d}}{\mathrm{d}r}\frac{\psi(r)}{r^{\nu}}
	\right]_{r=0} \nonumber\\
&= 	\left[
\frac{\varphi(r)}{r^{\nu}} + iL_{0}^{1 - 2\nu}r^{2\nu}
\frac{\mathrm{d}}{\mathrm{d}r}\frac{\varphi(r)}{r^{\nu}}
\right]_{r=0}^{\ast}
\left[
\frac{\psi(r)}{r^{\nu}} + iL_{0}^{1 - 2\nu}r^{2\nu}
\frac{\mathrm{d}}{\mathrm{d}r}\frac{\psi(r)}{r^{\nu}}
\right]_{r=0}, \label{eq:6}
\end{align}
where $L_{0}$ is a non-vanishing real length parameter, which is just introduced to adjust the length dimension of the equation.

The self-adjointness of $H_{\text{bulk}}$, however, further requires that $\varphi$ and $\psi$ obey the same boundary condition at the origin.
To this end we simply put $\varphi = \psi$ in \eqref{eq:6}, which reduces to the following equation:
\begin{align}
\left|
\frac{\psi(r)}{r^{\nu}} - iL_{0}^{1 - 2\nu}r^{2\nu}
\frac{\mathrm{d}}{\mathrm{d}r}\frac{\psi(r)}{r^{\nu}}
\right|_{r=0}^{2}
&= 	\left|
	\frac{\psi(r)}{r^{\nu}} + iL_{0}^{1 - 2\nu}r^{2\nu}
	\frac{\mathrm{d}}{\mathrm{d}r}\frac{\psi(r)}{r^{\nu}}
	\right|_{r=0}^{2}. \label{eq:7}
\end{align}
This equation shows that the complex number $\left[\frac{\psi(r)}{r^{\nu}} - iL_{0}^{1 - 2\nu}r^{2\nu}\frac{\mathrm{d}}{\mathrm{d}r}\frac{\psi(r)}{r^{\nu}}\right]_{r=0}$ has the same length as $\left[\frac{\psi(r)}{r^{\nu}} + iL_{0}^{1 - 2\nu}r^{2\nu}\frac{\mathrm{d}}{\mathrm{d}r}\frac{\psi(r)}{r^{\nu}}\right]_{r=0}$, which implies that these two complex numbers must be related by $U(1)$ transformation.
Thus we see that the whole self-adjoint domain of $H_{\text{bulk}}$ for $g > -1/4$ is specified by the following $U(1)$ family of boundary conditions:
\begin{align}
\left[
\frac{\psi(r)}{r^{\nu}} - iL_{0}^{1 - 2\nu}r^{2\nu}
\frac{\mathrm{d}}{\mathrm{d}r}\frac{\psi(r)}{r^{\nu}}
\right]_{r=0}
&= 	\mathrm{e}^{i\alpha}
	\left[
	\frac{\psi(r)}{r^{\nu}} + iL_{0}^{1 - 2\nu}r^{2\nu}
	\frac{\mathrm{d}}{\mathrm{d}r}\frac{\psi(r)}{r^{\nu}}
	\right]_{r=0}, \label{eq:8}
\end{align}
or, equivalently,
\begin{align}
\left[
\frac{\psi(r)}{r^{\nu}} + L_{0}^{1 - 2\nu}\left(\cot\frac{\alpha}{2}\right)
r^{2\nu}\frac{\mathrm{d}}{\mathrm{d}r}\frac{\psi(r)}{r^{\nu}}
\right]_{r=0}
&= 	0, \label{eq:9}
\end{align}
where $0\leq \alpha <2\pi$.
It should be emphasized that when we choose $\nu = \nu_{-}$ the boundary condition \eqref{eq:9} becomes well-defined because $\psi(r)/r^{\nu_{-}}$ has a definite boundary value.
Thus, by putting $\nu = \nu_{-}$ and using the identity $1 - 2\nu_{-} = \nu_{+} - \nu_{-}$ we finally obtain the boundary condition \eqref{eq:5.4}.

\end{document}